\documentstyle[11pt,twoside,fleqn]{article}   
\begin{document}

\setlength{\baselineskip}{5ex}


\newcommand{\ie}{{\it i.\,e.\ }}        \newcommand{\cf}{{\it cf.\ }}
\newcommand{\eg}{{\it e.\,g.\ }}
\newcommand{\etal}{{\it et al.}}

\newcommand{\fns}{\footnotesize}         \newcommand{\fnt}{\footnote}
\newcommand{\bect}{\begin{center}}
\newcommand{\enct}{\end{center}}
\newcommand{\befr}{\begin{flushright}}   \newcommand{\enfr}{\end{flushright}}
\newcommand{\befl}{\begin{flushleft}}    \newcommand{\enfl}{\end{flushleft}}
\newcommand{\beeq}{\begin{equation}}     \newcommand{\eneq}{\end{equation}}
\newcommand{\bear}{\begin{array}}        \newcommand{\enar}{\end{array}}
\newcommand{\bea}{\begin{eqnarray}}      \newcommand{\eea}{\end{eqnarray}}
\newcommand{\nid}{\noindent}           \newcommand{\hspi}{\hspace*{\parindent}}
\newcommand{\bsk}{\bigskip}
\newcommand{\msk}{\medskip}              \newcommand{\ssk}{\smallskip}

\def\go{ \mathrel{\raise.3ex\hbox{$>$}\mkern-14mu\lower0.6ex\hbox{$\sim$}} }
\def\lo{ \mathrel{\raise.3ex\hbox{$<$}\mkern-14mu\lower0.6ex\hbox{$\sim$}} }

\newcommand{\al}{\mbox{$\alpha$}}
\newcommand{\aJ}{\mbox{$\alpha_{J}$}}    \newcommand{\aQ}{\mbox{$\alpha_{Q}$}}
\newcommand{\hJ}{\mbox{$h_{J}$}}         \newcommand{\hQ}{\mbox{$h_{Q}$}}
\newcommand{\gm}{\mbox{$\gamma$}}        \newcommand{\Gm}{\mbox{$\Gamma$}}
\newcommand{\om}{\mbox{$\omega$}}        \newcommand{\Om}{\mbox{$\Omega$}}
\newcommand{\OmH}{\mbox{$\Omega_{H}$}}
\newcommand{\OmF}{\mbox{$\Omega_{F}$}}   \newcommand{\nb}{\mbox{$\nabla$}}
\newcommand{\vro}{\varrho}
\newcommand{\vre}{\mbox{$\varrho_{e}$}}
\newcommand{\vrm}{\mbox{$\varrho_{m}$}}\newcommand{\vp}{\mbox{$\varpi$}}
\newcommand{\th}{\mbox{$\theta$}}
\newcommand{\vh}{\mbox{$\varphi$}}       \newcommand{\ld}{\mbox{$\lambda$}}
\newcommand{\zt}{\mbox{$\zeta $}}        \newcommand{\bt}{\mbox{$\beta$}}
\newcommand{\dl}{\mbox{$\delta$}}        \newcommand{\Dl}{\mbox{$\Delta$}}
\newcommand{\sg}{\mbox{$\sigma$}}        \newcommand{\Sg}{\mbox{$\Sigma$}}
\newcommand{\ep}{\mbox{$\epsilon$}}
\newcommand{\tlbt}{\tilde{\bt}}          
\newcommand{\tlE}{\tilde{E}   }          \newcommand{\tlS}{\tilde{S}}
\newcommand{\tlu}{\tilde{u}}

\newcommand{\hspc}{\hspace*{1cm}}
\newcommand{\beqn}{\\ \msk $ \hspc}
\newcommand{\eeqn}[1]{\hfill (#1) $ \\ \msk}

\newcommand{\MC}{microcanonical} \newcommand{\Cnn}{canonical}

\newcommand{\Lbr}[1]{\left[#1\right]}   
\newcommand{\Lprth}[1]{\left(#1\right)} 
\newcommand{\dr}[2]{\frac{d#1}{d#2}}    
\newcommand{\drh}[3]{\frac{d^{#1}#2}{d#3^{#1}}} 
\newcommand{\pldr}[2]{\frac{\pl#1}{\pl#2}}  
\newcommand{\pldrh}[3]{\frac{\pl^{#1}#2}{\pl#3^{#1}}}  

\newcommand{\msf}{mean square fluctuation}
\newcommand{\MSF}[1]{\overline{(\Dl #1)^2}} 
\newcommand{\MSFbt}{\MSF{\bt}}          
\newcommand{\MSFE}{\MSF{E}}             

\newcommand{\vep}[1]{\mbox{$\varepsilon_{#1}$}}  
\newcommand{\vc}[1]{\mbox{\boldmath $#1$}}       
\newcommand{\ra}{\mbox{$\rightarrow$}}   \newcommand{\pl}{\partial}

\newcommand{\dps}{\displaystyle}
\newcommand{\fracdps}[2]{\frac{\dps #1}{\dps #2}}
\newcommand{\drdps}[2]{\frac{\dps d#1}{\dps d#2}}    
\newcommand{\drhdps}[3]{\frac{\dps d^{#1}#2}{\dps d#3^{#1}}}
\newcommand{\pldrdps}[2]{\frac{\dps\pl#1}{\dps\pl#2}}  
\newcommand{\pldrhdps}[3]{\frac{\dps\pl^{#1}#2}{\dps\pl#3^{#1}}}

\newcommand{\Schw}{Schwarzschild}        \newcommand{\KN}{Kerr-Newman}
\newcommand{\Pc}{Poincar\'e}             \newcommand{\bh}{black hole}
\newcommand{\tc}{thermodynamic coefficient}
\newcommand{\Ecal}{{\cal E}}             \newcommand{\Bcal}{b}
\newcommand{\Jcal}{{\cal J}}             \newcommand{\Ical}{{\ell}}
\newcommand{\Dcal}{{\cal D}}             \newcommand{\Pcal}{{\cal P}}
\newcommand{\Ccal}{{\cal C}}
\newcommand{\btrad}{\bt_{rad}}           \newcommand{\btbh}{\bt_{bh}}
\newcommand{\Ebh}{E_{bh}}                \newcommand{\Erad}{E_{rad}}
\newcommand{\EbhC}{\Ecal_{{bh}}}         \newcommand{\EradC}{\Ecal_{{rad}}}
\newcommand{\Sbh}{S_{{bh}}}              \newcommand{\Srad}{S_{rad}}
\newcommand{\Sein}{S_{{Ein}}}            \newcommand{\Crad}{C_{rad}}
\newcommand{\gmunu}{g_{\mu\nu}}
\newcommand{\Dltlu}{\Dl\tilde{u}}        \newcommand{\Dlu}{\Dl u}

\newcommand{\NV}[2]{#1\cdot 10^{#2}}   
\newcommand{\vol}[1]{{\bf #1}}     
\newcommand{\ibid}{{\it ibid}}
\newcommand{\ApJ}{{\it Astrophys.\,J.}}
\newcommand{\CQG}{{\it Class.\,Quantum Grav.}}
\newcommand{\MN}{{\it Mon.\,Not.\,R.\,Astron.\,Soc.}}
\newcommand{\PRD}{{\it Phys.\,Rev.\,}D}

\newcommand{\btprm}{\bt^{\prime}}  \newcommand{\Eprm}{E^{\prime}}
\newcommand{\Vprm}{V^{\prime}}
\newcommand{\EQ}{equation}


\vspace*{2cm}
\centerline{\LARGE {\bf Thermodynamics of a black hole in a cavity }}
\msk
\bsk
\bsk
\bsk
\bsk
\renewcommand{\thefootnote}{\fnsymbol{footnote}}
\footnotetext[4]{Present address: Laboratoire de Physique Theorique de l' Ecole
Normale Superieure, 24 Rue Lhomond, 75005 Paris, France (E-mail:
parenta@physique.ens.fr)}
\footnotetext[2]{Permanent address: The Racah Institute of Physics,
   Jerusalem 91904, Israel (E-mail: jkatz@hujivms.bitnet)}
\befr \parbox[t]{14.4cm}{ %
  {\large\bf Renaud Parentani$^1$\footnotemark[4],
       Joseph Katz$^2$\footnotemark[2] and Isao Okamoto$^2$} \\[8mm]
$^1$ The Racah Institute of Physics, The Hebrew University of
Jerusalem, Jerusalem 91904, Israel  \\[1mm]
$^2$ Division of Theoretical Astrophysics, National Astronomical Observatory,
     Mizusawa, Iwate 023, Japan (E-mail: okamoto@gprx.miz.nao.ac.jp) \\[5mm]
   } %
\enfr
\vspace*{1cm}
\vspace*{1cm}
\vspace*{1cm}
\vspace*{5mm}
\befr\parbox[t]{16.4cm}{  %
{\bf Abstract. }
We present a unified thermodynamical description of the configurations
consisting on self-gravitating radiation with or without a black hole. We
compute the thermal fluctuations and evaluate where will they induce a
transition from metastable configurations towards stable ones. We show that
the probability of finding such a transition is exponentially small. This
indicates that, in a sequence of quasi equilibrium configurations, the system
will remain in the metastable states till it approaches very closely the
critical point beyond which no metastable configuration exists. Near that
point, we relate the divergence of the local temperature fluctuations to the
approach of the instability of the whole system, thereby generalizing the
usual fluctuations analysis in the cases where long range forces are present.
When angular momentum is added to the cavity, the above picture is slightly
modified. Nevertheless, at high angular momentum, the black hole loses most of
its mass before it reaches the critical point at which it evaporates
completely.
}  
\enfr
\vspace*{1cm}

\newpage
\befl
{\bf 1.\ Introduction and summary} \\[2mm]
\enfl
A black hole in an isolated cavity which is bigger than its Schwarzschild
radius can be in equilibrium with surrounding radiation only if the total
energy $E$ of the system is greater than a critical value $E_C$, which
depends on the volume of the cavity and the number of fields in the radiation
(Hawking 1976). While local equilibrium configurations exist for energies
greater than $E_C$, it is generally assumed that black holes must evaporate
if $E_C<E<E_B\simeq 1.3E_C$ since for $E<E_B$ pure radiation has more entropy
than the composite system (Gibbons and Perry 1978).

This holds only if the system is given unlimited time to relax. If one
considers instead a sequence of configurations starting, say, from an
equilibrium at $E=E_B$ and in which the total energy is decreased
quasi-statically, the system reaches new local equilibrium configurations in
a finite time (Zurek 1980). During that time, the probability of finding a
fluctuation high enough that the black hole completely evaporates is
exponentially small. So small that the black hole is most likely to survive
in a metastable superheated state up to energies very close indeed to $E_C$
(Okamoto, Katz and Parentani 1994). This is true for cavities whose radius
$L^*$ is greater than $10^6$ Planck lengths $l_P=(\hbar G/c^3)^{1/2}\approx
10^{-33}$cm ($L\equiv L^*/l_P>10^6$). For those cavities, back-reaction due
to quantum matter effects (York 1985), quantum gravity and spontaneous
nucleation due to thermal fluctuations (Piran and Wald 1982) may be neglected
since the maximum temperature, reached by equilibrium configurations, is
always smaller than $T_D \simeq 0.37L^{-1/2}$\fnt{ 
Throughout the text numerical values are given with two significant digits
}   
(in Planck units). Thus more than three orders of magnitude separate the
equilibrium temperature from the Planck one. This indicates that a
thermodynamic analysis may be safely performed. On the contrary, for smaller
boxes the validity of this analysis becomes dubious.

Self-gravitating thermal radiation exhibits a similar behavior in an
evolutionary sequence in which the total energy increases quasi-statically.
Indeed, there is now an upper limit of energy $E_A \simeq 0.25L$ (here $E$
is measured in Planck units) above which no equilibrium configuration exists
(Klein 1947). Thus for energies $E_B<E<E_A$ ($E_B \simeq 0.20 L^{3/5}\ll E_A
\simeq 0.25L$ for $L>10^6$) pure radiation is in metastable superheated
states. We shall prove that, as in the black hole situation, pure radiation
is most likely to remain in those metastable states almost up to $E_A$. At
that point, a thermodynamic as well as dynamical instability (Sorkin, Wald and
Zhang 1980) develops and most of the radiation collapses into a black hole
leading to a composite state of a black hole in equilibrium with surrounding
left-over radiation. The spontaneous evaporation of the black hole into pure
radiation near $E_C$ and the collapse of the radiation near $E_A$ provides
thus a unified picture relating the different phases of a black hole and
self-gravitating radiation in a cavity.

Uniform rotation added to the cavity gives a total angular momentum $J$ shared
between the radiation and the black hole. The equilibrium configurations
change in the following way. The critical energy $E_C(J)$ at which the black
hole evaporates is now a function of $J$. It increases with increasing angular
momentum while the fraction of black hole energy at that point $E_{bhC}/E_C$
decreases from $4/5$ at $J=0$ to almost
zero in the limiting rotating case: $J=L^2/2$.
Hence for fast rotating cavities, at fixed $J$ and with slowly decreasing
total energy, superheated Kerr black holes will evaporate with little latent
heat, since black holes survive almost down to the critical point $E_C(J)$.

All of this concerns microcanonical situations in which the total energy is
the control parameter. In canonical situations, with the temperature kept
fixed, almost all the configurations wherein a black hole coexists with
radiation are unstable. There exits nevertheless a narrow range of energies
for very fast rotating cavities in which the canonical ensemble is stable.
One thus recovers in this narrow range the already noticed flip of the heat
capacity for Kerr black holes (Davies 1981, Kaburaki \etal\ 1993; \cf Katz
\etal\ 1993 for Kerr-Newman black holes). When the radiation is included, this
happens, however, when gravitational effects (neglected in the present
analysis) are important.

In this paper, we analyze the stability limits, the different degrees of
instability and the fluctuations of each of the two states of the system
(with or without black hole) in evolutionary sequences of equilibrium
configurations, with and without angular momentum. The analysis is purely
thermodynamic. We assume there exists a state function whose stationary value
is a function of the total mass-energy $E$, total angular momentum $J$ and
the volume of the cavity $V=\ld L^3$. We consider separately the
microcanonical and the canonical ensembles since they are not equivalent.
In a microcanonical ensemble, $E$ is the main control parameter and the
inverse temperature
   \beeq 
   \frac{1}{kT} =\frac{\pl (S/k)}{\pl E}=\bt(E,J,L)
   \label{eq:S1.1}
   \eneq
is the conjugate parameter of $E$ with respect to the total entropy $S$. In
a canonical ensemble, the Massieu function is not $S$ but $S/k-\bt E= -\bt F$
($F$ is the free energy), the main control parameter is $\bt$ and the
conjugate parameter of $\bt$ with respect to $-\bt F$ is
   \beeq 
   \frac{\pl(-\bt F)}{\pl\bt} = -E(\bt,J,L).
   \label{eq:S12}
   \eneq
The linear series of equilibrium configurations in both ensembles, $\bt(E)$
and $-E(\bt)$ at fixed $J$ and $L$, are identical but the stability limits
are not. Stability limits are obtained by applying the \Pc\ method to linear
series (see Ledoux 1958). We shall consider the linear series of conjugate
parameters (Katz 1978, 1979; Thompson 1979) since this is the most appropriate
way to observe changes of stability.

While state functions out of equilibrium, $\Om$, are never used, some
assumptions must be made about $\Om$ which are stated in section 2 where the
extremization of the entropy is related to the Einstein equations when
self-gravity is taken into account. Stability analysis in terms of \Pc's
method is then briefly reviewed. We shall see that we need to know $\bt(E)$
only and not $\Om$ itself, a useful feature when analyzing the stability of
self-bound radiation in a box. Furthermore, having determined $\bt(E)$ from the
solution of the Einstein equations, we do not need to attribute {\it a priori}
an entropy to the black hole. Instead, we recover it from the equilibrium
configurations.

The theory of fluctuations is then considered. Fluctuation theory as given by
Landau and Lifshitz (1980) and Callen (1985) is not entirely applicable to
self-gravitating systems for the reason that when gravitational interactions
are important one cannot anymore split the system into a little subsystem and
the rest which behaves like a heat reservoir for the little one. Nevertheless,
we shall see that the mean quadratic fluctuations of the inverse temperature,
in a small subsystem, are simply related to its own thermodynamic parameters
as well as the parameters of the whole system.

Section 4 gives details about $\bt(E)$ in non-rotating and rotating cavities.
Fluctuations near turning points are calculated in section 5 and a summary of
stability conditions and phase transitions in slowly evolving systems through
a succession of quasi-equilibrium states is described in section 6.  \\

\befl
{\bf 2.\ The stability of equilibrium configurations in mean field
         theory} \\[2mm]
\enfl
In a stationary axially symmetric distribution of matter in local
thermodynamic equilibrium, with fixed total mass-energy and angular momentum,
and with Einstein constraint equations given, the total entropy $S$ is
stationary, $\dl S=0$, if Einstein ``dynamical" equations are satisfied.
Then Tolman's (1934) thermodynamic equilibrium conditions for local
temperature hold and the angular velocity is uniform [Katz and Manor (1975);
for non-rotating, spherical configurations see Cocke (1965)]. Here we shall
use the partition function (Hawking 1978, Horwitz and Weil 1982, York 1988 and
Brown \etal\ 1990) to sketch a formal deduction of both equilibrium and
stability conditions for spherical distributions with zero angular momentum.
This procedure will naturally exhibit the relations between the Einstein
equations, quantum field theory in curved space and the extremization of the
entropy.

Consider, for definiteness, the state function $\Om$ which encodes the total
number of states of a scalar field in a curved spacetime with total
mass-energy $E$ as measured from infinity. The field is confined to a
spherical box of ``radius" $L$ (\ie\ the surface of the cavity is $4 \pi
L^2$). Then $\Om(E,L)$ is
     \beeq  
      \Om(E,L)=e^{S(E,L)} =Tr\dl(E-H)
      =\int\frac{dt}{2\pi}e^{iEt}
      \int\frac{\Dcal \gmunu}{{\cal D}iff}\Dcal\phi\,
      \exp\Lbr{i\Sein+iS_{\phi}}.    \label{eq:S2.1}
     \eneq
The denominator ``${\cal D}iff$" indicates that one should not integrate
over geometries related by a diffeomorphism; $\Sein=\int d^4 x \sqrt{-g}R$,
and $S_{\phi}=\int d^4 x\sqrt{-g}g^{\mu\nu}\pl_{\mu}\phi\pl_{\nu}\phi$. Since
there is no unique definition of a local matter Hamiltonian in general
relativity, we have used the path integral formalism. We refer to Regge and
Teitelboim (1974) for a definition of energy in asymptotically flat
spacetimes. Integrating in (\ref{eq:S2.1}) over all periodic matter
configurations (of period $t$), one obtains
   \beeq	
   \Om(E,L) =\int\frac{dt}{2\pi} \int\frac{\Dcal \gmunu}{{\cal D}iff}
     \exp\Lbr{iEt+i\Sein}\,Z[\{\gmunu\};it]
   \label{eq:S2.2}
   \eneq
where $Z$ is the partition function at fixed Lorentzian time $t$ in the
background geometry $\{\gmunu \}$.

Calculating (\ref{eq:S2.2}) at the stationary configuration of $iEt+i\Sein+
\ln Z[\{\gmunu\};it]$ gives the dominant contribution to $\Om$. At the
stationary point, $\gmunu$ satisfies the time independent Einstein equations
with an energy tensor of a thermal bath if at the saddle point $t=-i\bt$. In
the absence of horizons, the matter field configurations are defined for
$0\leq r\leq L$ and one recovers, up to quantum matter corrections, the \EQ s
of a self-gravitating perfect fluid (Klein 1947, Gibbons and Hawking 1978). In
the presence of horizons (see Carlip and Teitelboim 1993 for the treatment of
the boundary term at the horizon), the regularity of the stress energy-momentum
tensor, needed to satisfy Einstein equations, fixes the temperature of the
matter (Hawking 1978). Then by integrating the time-time Einstein constraint
\EQ\ one obtains
   \beeq  
   m(r,M_{bh})=M_{bh}+ 4 \pi \int^r_{2M_{bh}}dr'r'^2\langle T^t_t (r')
                     \rangle_{HH,M_{bh}}
   \label{eq:S2.3}
   \eneq
where $\langle T^t_t\rangle_{HH,M_{bh}}$ is the expectation value of the
energy density, outside a black hole of mass $M_{bh}$, in the
Hartle-Hawking vacuum (Howard 1984). Since $E=m(L,M_{bh})$, one may invert
this relation and express $M_{bh}$ in terms of $E$ and $L$. Therefore, one
obtains the sought-for $\bt=\bt(E)$ law (with $L$ held fixed). Notice that
Einstein equations with back-reaction taken into account provide an
alternative way to determine the entropy of the black hole. Indeed, by
integrating $\bt(E)dE$ one obtains the total entropy $S(E,L)$, which reduces
to the black hole entropy when the radiation energy is negligible. One may
also subtract from $S(E,L)$ the entropy of the radiation, but there is an
arbitrariness in this subtraction \,---\,since there is no unique definition
of the entropy density nearby the hole where the Tolman relation breaks
down\,---\,which indicates that the concept of the entropy of an isolated
black hole is probably meaningless.

In order to illustrate the \Pc\ method, we perform the $t$-integration in
(\ref{eq:S2.2}) (Horwitz and Katz 1978) which gives an $\Om$ of the form
   \beeq	
   \Om(E,L)=\int \frac{\Dcal \gmunu}{{\cal D}iff} e^{ -w(E,L,\{\gmunu\}) }.
   \label{eq:S2.4}
   \eneq
We do not need to be specific about the function $w(E,L,\{\gmunu\})$
precisely because \Pc\ analysis deals with the succession of extrema of
$w(E,L,\{\gmunu\})$ and not the function itself. A specific example is given
in Sorkin \etal\ (1980) in the case of pure radiation without horizon, wherein
$w$ depends on $g_{rr}$ only.

The stationary value of $w$ is the classical equilibrium entropy $S$. The
local stability of equilibrium configurations is controlled by the quadratic
fluctuations of $w$. Imagine we make a Fourier decomposition of the $\gmunu$'s.
Since the domain of existence is finite $(0\leq r\leq L)$, the $\gmunu$'s are
replaced by a denumerable set of variables, say, $x^i$ ($i=1,2,....$). To
order two, $w$ may thus be written
  \beeq	 
  w \approx S+\frac12 \Lprth{\frac{\pl^2 w}{\pl x^i \pl x^j}}_{e}
              (x^i-X^i)(x^j-X^j)
  \label{eq:S2.5}
  \eneq
where an index $e$ means `at equilibrium' [thus $S=w_e$] and $X^i(E,L)$ are
equilibrium values of $x^i$. Let $\ld_i$'s denote the (ordered: $\ld_1\leq
\ld_2 \leq ...$) elements of the diagonalized matrix $-(\pl^2 w/\pl x^i\pl
x^j)$. The $\ld_i$'s are known as the \Pc\ coefficients of stability (Lyttleton
 1953). The equilibrium is thermodynamically (locally) stable, or stable for
arbitrary small fluctuations, if and only if all the $\ld_i$'s are positive
\ie\ if $\ld_1 > 0$. In unstable situations, the number of $\ld_i<0$
characterize the degree of instability.

The matrix $-(\pl^2 w/\pl x^i\pl x^j)_e$ is a second order differential
operator of dimension one over a finite domain. We may assume that the spectrum
of eigenvalues is non-degenerate: $\ld_1<\ld_2<...$\,. Indeed, the slightest
asymetric perturbation in a system would lift the degeneracy (Thompson and Hut
1973). Having then a non-degenerate spectrum  of \Pc\ coefficients, the
following properties hold (Katz 1978, 1979): \\[2mm]
(a) Consider the linear series $\bt(E)$. Changes of stability along the
linear series can only occur at vertical tangents, $\pl\bt/\pl E=\pm\infty$,
like point A or C in figure 1. Such points where $E$ is locally minimum or
maximum are called {\em turning points}. Between two vertical tangents all
equilibrium configurations have the same degree of instability. \\[1mm]
(b) In the neighborhood of a vertical tangent, when the linear series turns
clockwise (its tangent goes from negative to positive values through
infinity), one \Pc\ coefficient changes sign from negative to positive value.
That is, the system becomes stable or less  unstable. If the linear series
turns counter-clockwise, the changes of sign are reversed and the system
becomes unstable or more unstable. It is thus enough to know the degree of
stability of one configuration to know the degree of stability of all
configurations. \\[1mm]
(c) Of particular interest are linear series with multiple turning points
spiraling inwards against the clock. If we follow the spiral towards the limit
point, we meet a succession of vertical tangents and beyond each one, an
additional \Pc\ coefficient becomes negative. Counter-clockwise spirals
represent thus a succession of equilibrium configurations that are more and
more unstable. \\[1mm]
Upon considering different ensembles, the following properties hold (Parentani
1994): \\
(d) The most stable ensemble is always the most isolated one. The degree of
stability of equilibrium configurations in any ensemble related by a Legendre
transformation (which expresses the contact with a reservoir) to the most
stable is immediately
known if one knows the degree of stability in the most stable
ensemble. \\

\befl
{\bf 3.\ Fluctuations in gravitating systems }
\enfl
In classical thermodynamics, it is well known (Landau and Lifshitz 1980 (\S
112), Callen 1985 and Landsberg 1990) that the mean square fluctuations of the
fundamental thermodynamic quantities pertaining to any small part of a system
(or to the system as a whole) are related to the specific heat. For instance,
in a \Cnn\ ensemble of total volume $V$, the mean quadratic fluctuations
$\MSFE$ of the energy $E$ induced by the contact with the reservoir are given
by
   \beeq
   \MSFE=\pldr{(-E)}{\bt}=\frac{C_V}{\bt^2}
   \label{eq:S3.1}
   \eneq 
where $C_V$ is the heat capacity at constant volume $V$. The \msf s $\MSFbt$
of the inverse temperature which result from those energy fluctuations are
given by
  \beeq
  \MSFbt=-\pldr{\bt}{E}= \frac{\bt^2}{C_V}.
  \label{eq:S3.2}
  \eneq  
since $\bt$ is a function of $E$ only.

In a \MC\ ensemble, these fluctuations
vanish since the total energy $E$ is kept fixed. Nevertheless, within a small
subsystem of volume $V'$, the temperature fluctuates and the \msf s of $\bt'$
are given by
\beeq
  \MSF{\bt'}=-\pldr{\bt'}{E'}= \frac{\bt'^2}{C_{V'}}.
  \label{eq:S3.3}
  \eneq  
where $C_{V'}$ is the specific heat of the subsystem and $E'$ its energy. This
equation is valid only if the specific heat of the rest of  the system is much
bigger that the one of the subsystem (for homogeneous systems it requires
$\Vprm\ll V$). The equivalence of ensembles (the equality of the fluctuations
(\ref{eq:S3.3}) whether one works in the \MC\ or the \Cnn\ ensemble) means
therefore that the rest of the system can be correctly treated as a heat
reservoir for the little subsystem. Finally, we recall that the 'true'
fluctuations are given by these estimates only if the characteristic dynamical
time for the fluctuations to evolve is much bigger than $\bt$ itself.

In gravitating systems, there are long range forces. Therefore, when
gravitational interactions are important, one cannot treat the rest of the
system as a passive reservoir. Indeed the existence of stable \MC\ ensembles
with negative specific heat indicates the decisive role of the energy
constraint between the little subsystem and the rest. Furthermore, we stress
the fact that when a \MC\ ensemble {\em approaches instability}, its heat
capacity given in \EQ\ (\ref{eq:S3.1}) is always {\it negative} (thus the
\Cnn\ ensemble is already unstable) for stable states and {\it positive} in
unstable states (for which the \Cnn\ ensemble is still unstable)
see section 4 and, for instance, figure 1.

The relation  between mean square fluctuations and thermodynamic coefficients
in gravitating systems is thus more complicated than \EQ\ (\ref{eq:S3.3})
and we shall display it in two steps. We shall first see that, since the
vicinity of a turning point is dominated by a single eigenmode, one can relate
the fluctuations of the least stable mode $x^1$ to the heat capacity of the
whole system $C_V$. Then, we shall relate the fluctuations of the inverse
temperature within a small subsystem to $C_V$ itself. To prove the first point,
consider a \MC\ ensemble (with $J=0$) in which we define a {\em temperature
function} $\tlbt(x^i; E,V)$ away from equilibrium (see also Okamoto \etal\
1994)
  \beeq
  \tlbt=\pldr{w}{E}.
  \label{eq:S3.4}  
  \eneq
At equilibrium one has $\tlbt (X^i(E,V);E,V)=\bt (E,V)$ where the equilibrium
values $x^i= X^i(E,V)$ are solutions of
  \beeq
  \pldr{w}{x^i}=0.
  \label{eq:S3.5}  
  \eneq
Therefore, the slope of the linear series and the derivatives of $\tlbt$ are
related by
  \beeq
  \pldr{\bt(E,V)}{E}=\Lprth{\pldr{\tlbt(x^i;E,V)}{E}}_e
                   + \sum_i \Lprth{\pldr{\tlbt}{x^i}}_e \pldr{X^i(E,V)}{E}
  \label{eq:S3.6}  
  \eneq
The derivative $\pl X^i/\pl E$ can be obtained from equation (\ref{eq:S3.5})
by the following identity
  \beeq
  \Lprth{ \pldr{ }{E} \Lbr{ \Lprth{\pldr{w}{x^i}}_e }}
  = \Lprth{\pldr{\tlbt}{x^i}}_e-\ld_i\pldr{X^i}{E}\equiv 0
  \label{eq:S3.7}  
  \eneq
where there is no summation on $i$. Thus, from \EQ\ (\ref{eq:S3.7}) we deduce
$\pl X^i/\pl E$ in terms of $(\pl\tlbt/\pl x^i)_e$ which we may insert into
(\ref{eq:S3.6}) to obtain
  \beeq
  \pldr{\bt}{E}=\Lprth{ \pldr{\tlbt}{E} }_e
                +\sum_i \frac{ (\pl\tlbt/\pl x^i)_{e}^2}{\ld_i} .
  \label{eq:S3.8} 
  \eneq
Consider now a linear series of stable configurations near a turning point.
At that point the first eigenvalue $\ld_1$ changes sign. As a result the right
hand side of \EQ\ (\ref{eq:S3.8}) is entirely dominated by the first term
  \beeq
  \pldr{\bt}{E}
  \simeq \frac{1}{\ld_1}\Lprth{\pldr{\tlbt}{x^1}}_e^2.
  \label{eq:S3.9} 
  \eneq
The changes of $w$ are also dominated by the fluctuations of $x^1$ and given by
   \beeq
   w\simeq S-\frac12 \ld_1 (\Dl x^1)^2
   \label{eq:S3.10} 
   \eneq
which, with equation (\ref{eq:S3.9}), becomes
   \beeq
   w\simeq S-\frac12 \frac{(\Dl\tlbt)^2}{\pl\bt/\pl E}
   \label{eq:S3.11} 
   \eneq
where
   \beeq
   \Dl\tlbt=\Lprth{\pldr{\tlbt}{x^1}}_e \Dl x^1.
   \label{eq:S3.12}  
   \eneq
This $\Dl\tlbt$ represents fluctuations in the inverse-temperature {\em
function} induced, near the critical point, by the fluctuations of the least
stable mode $x^1$.  We can now use the standard arguments of fluctuation
theory (Landau and Lifshitz 1980) and say that the probability $dW$ for a
fluctuation of $\tlbt$ in the range $\tlbt+\Dl\tlbt$ and $\tlbt+\Dl\tlbt+
d\tlbt$ is proportional to $\exp(w-S)$ and therefore,
   \beeq
   dW=\frac{1}{\sqrt{2\pi}}\frac{d\tlbt}{\sqrt{(\pl\bt/\pl E)}}
           \exp\Lbr{ -\frac12 \frac{(\Dl\tlbt)^2}{(\pl\bt/\pl E)} }
   \label{eq:S3.13} 
   \eneq
Thus the mean square fluctuations of $\tlbt$ are given by
   \beeq
   \MSF{\tlbt}=\pldr{\bt}{E}=-\frac{\bt^2}{C_V}.
   \label{eq:S3.14}  
   \eneq
They are bounded when the \MC\ ensemble is stable (\ie when $\ld_1>0 $ or
as we already point out when $C_V<0$.
See \EQ\ (\ref{eq:S3.9})). It is convenient to express $dW$ in terms
of $C_V$ and of the dimensionless ratio
   \beeq
   \frac{\tlbt(x^i;E,L)}{\bt}\equiv \tlu.
   \label{eq:S3.141} 
   \eneq
Then
   \beeq
   dW=\sqrt{\frac{-C_V}{2\pi}}\,\exp \Lbr{+\frac12 C_V(\Dl\tlu)^2} d\tlu
   \label{eq:S3.142} 
   \eneq
and, following (\ref{eq:S3.14}) the mean square fluctuations of $\tlu$ are
   \beeq
   \MSF{\tlu}=-C_V^{-1}.
   \label{eq:S3.143} \eneq  
Our task now is to relate those rather formal fluctuations of $\tlbt$ to the
fluctuations of the temperature within a small part of the system as well as
to understand the origin of the
unusual sign in equation (\ref{eq:S3.14}). As we have already
said, in the presence of long range forces, one cannot exactly split a system
into two parts since there is no more a local definition of energy. If
nevertheless some small part is less coupled gravitationally to the rest of
the system, one may use it as the "small" system in which one can compute the
fluctuations of $\bt$. When this is not the case, we shall see that one can
still consider the fluctuations within the outermost layer of radiation even
though the energy into that layer is not well defined. We designate by the
subscript $2$ the little subsystem which is confined for radii $L-l<r<L$
(where $l\ll L$). The subscript $1$ refers to the rest of the system which is
thus either the black hole surrounded by radiation up to that last layer, or
pure radiation. Instead of the energy, we shall use the \Schw\ mass (since it
is a local quantity) and we thus split it into $E_1$ and $E-E_1$. We shall
see that the partition mass $E_1$ will play a role very similar to the least
stable mode $x^1$. This is because $\pl_{x^1}E_1\neq 0$.

Let $w(E_1;E,V)$ be the entropy out of equilibrium by which we now mean that
the only variable out of equilibrium is $E_1$. All other variables have been
replaced by their equilibrium values. Then
   \beeq
   w(E_1;E,V)= S_1(E_1,V_1)+ S_2 (E-E_1,V-V_1;E_1)
   \label{eq:S3.15} 
   \eneq
where $S_1$ and $S_2$ are the entropies of the two parts. We emphasize the
double dependence of $E_1$ in $S_2$. When $E_1$ varies, the mass left over in
the system 2 is $E-E_1$ but the gravitational potential in which 2 evolves is
changed as well. This later dependence is nevertheless parametric if $l\ll L$.
By parametric we mean that upon taking derivatives with respect to $E_1$,
this later dependence gives an additional term which is  ${\cal O}(l/L)$
smaller than the usual term and which may be safely neglected. Equilibrium
between the two parts requires, as usual, the equality of the temperatures:
   \beeq
   \pl_{E_1}w(E_1;E,V)=\pl_{E_1}S_1+\pl_{E_1}S_2 \equiv \bt_1-\bt_2=0
   \label{eq:S3.16} 
   \eneq
where $\bt_1$ and $\bt_2$ are the inverse temperatures of the two parts and
where we have neglected the additional term (This does not mean that we neglect
completely the extra dependence in $E_1$, for instance, $\bt_2$ depends
explicitly on $E_1$ trough the Tolman dependence (see the Appendix \EQ\
(A.3))).
Then the lowest order variations of $w$ near equilibrium due to an exchange
of energy $\Dl E_1$ is:
   \beeq
   w-S = \frac12 (\pl_{E_1}^2 w) (\Dl E_1)^2
       = \frac12\Lbr{\pl_{E_1}\bt_{1}+\pl_{E_2}\bt_{2}}(\Dl E_2)^2  
   \eneq
This energy fluctuation induces, in turn, a fluctuation of $\bt_2$ in the
small part given by
   \beeq
   \Dl\bt_2=\Dl E_2 (\pl_{E_2}\bt_2)  
   \eneq
Then the entropy fluctuation induced by the latter one is
   \beeq
   w-S = \frac12 (\pl_{\bt_2} E_2 )
           \Lbr{1+(\pl_{E_1}\bt_1)(\pl_{\bt_2}E_2)}(\Dl\bt_2)^2
        =\frac12 (\pl_{\bt_2}E_2)\dr{E}{\bt}(\pl_{E_1}\bt_1)(\Dl\bt_2)^2 
   \label{eq:S3.29} \eneq
since the slope of the $\bt(E)$ curve of the entire system is given by
   \beeq
   \dr{\bt}{E} =(\pl_{E_2}\bt_2)\Lbr{1-\dr{E_1}{E} }
      = (\pl_{E_1}\bt_1) \Lbr{1+(\pl_{E_1}\bt_1)\pl_{\bt_2}E_{2}}^{-1} 
   \label{eq:S3.30}\eneq
because
   \beeq
   dE = dE_1 \Lbr{1+(\pl_{\bt_2}E_2)(\pl_{E_1}\bt_1)} 
   \eneq
Thus the mean square fluctuations of $\bt_2$ given by the inverse coefficient
appearing in \EQ\ (\ref{eq:S3.29}) are
   \beeq
   \MSF{\bt_2}=-(\pl_{E_2}\bt_2) \dr{\bt}{E}(\pl_{\bt_1}E_1). 
   \label{eq:S3.32} \eneq
This is the generalisation of \EQ\ (\ref{eq:S3.3}) when one cannot treat the
rest of the system as a passive reservoir. Indeed, one does recover
(\ref{eq:S3.3}) when the subsystem 1 can be approximated by the whole system
in which case the coefficient of $\pl_{E_2}\bt_2$ is 1. One sees also that the
point wherein the fluctuations of $\bt_2$ will diverge is controlled entirely
by the divergence of $d\bt/dE$ when the denominator in \EQ\ (\ref{eq:S3.30})
vanishes. That is, one probes locally, through the fluctuations in the small
subsystem the stability of the whole system $1+2$. Furthermore, near the
critical point, the mean fluctuations of $\bt_2$ are equal to the fluctuations
of $\tlbt$ (and therefore independent of $l$) since $(\pl_{\bt_2}E_2)
(\pl_{E_1}\bt_1)=-1$ at the critical point as seen in \EQ\ (\ref{eq:S3.30}).
This later factor of $-1$ explains the unusual sign in \EQ\ (\ref{eq:S3.14}).

For the interested reader, we also point out the analogy of \EQ\
(\ref{eq:S3.32}) which gives the fluctuations when the rest of the system
cannot be treated as a reservoir and the expression which relates the
fluctuations in one ensemble to the fluctuations in another ensemble related
to the first one by a Legendre transformation (Parentani 1994). In both cases
when the ensembles are nonequivalent, the fluctuations are controlled by the
correction factor: $\dr{\bt}{E}(\pl_{\bt_1}E_1)$,
the coefficient of $\pl_{E_2}\bt_2$ in \EQ\ (\ref{eq:S3.32}).
And it is only when the ensembles are equivalent that this factor reduces to 1.
Another common feature is the fact that when $d\bt/dE =0$ it does not imply
that the fluctuations of $\bt_2$ vanishes because $\pl_{E_1}\bt_1$ vanishes
as well, see \EQ\ (\ref{eq:S3.30}).

For a \Cnn\ ensemble, when one consider configurations which are stable
microcanonically, the necessary and sufficient condition for stability is the
positivity of $C_V$. For those configurations only one may safely apply the
analysis of Landau and Lifshitz and find that the fluctuations of the total
energy are indeed given by \EQ\ (\ref{eq:S3.1}). Finally let us mention that
equivalent ensembles would have vertical and horizontal slopes in $\bt(E)$ at
practically the same point since both ensembles become simultaneously unstable.
Thus the curve $\bt(E)$ would make an angle bigger than $90^{\circ}$. Nothing
like that is happening in the following situations.  \\

\befl
{\bf 4.\ The linear series of conjugate parameters } \\[2mm]
\enfl
In Planck units, $E$, $J$ and $\bt$ are respectively given by
   \beeq 
   E=\frac{E^{*}}{E_P}, \quad J=\frac{J^{*}}{\hbar}, \quad \bt=\bt^{*}E_P
   \label{eq:S4.1}
   \eneq
in which $E_P=\sqrt{\hbar c^5/G}\approx 10^{16}$erg and asterisks denote
quantities in cgs units. Instead of using Planck units for $E$, $J$ and $\bt$,
we shall use dimentionless quantities expressed
in terms of $G$, $c$ and $L^*$ as is common
in classical general relativity.
One thus define
   \beeq 
   \Ecal=\frac{GE^*}{c^4L^*}, \quad \Jcal=\frac{GJ^*}{c^3 L^{*\,2}},
   \quad \Bcal=\frac{\hbar c\bt^*}{8\pi L^*}.
   \label{eq:S4.2}
   \eneq
Then $\Ecal$ varies between $0$ and $1/2$ (the \Schw\ limit) and since $J^*<E^
*L^*/c$ (the centrifugal limit), $\Jcal$ is always smaller than $\Ecal$. Thus,
    \beeq 
    0\leq \Jcal\leq \Ecal\leq\frac12.
    \label{eq:S4.3}
    \eneq
We shall see nevertheless the appearance of a fourth dimensionless quantity
$L=L^*/l_P$ which will reflect the fact that the Hawking temperature has a
quantum origin. Finally, the relation between $E$, $J$, $\bt$ and $\Ecal$,
$\Jcal$, $\Bcal$ is
    \beeq 
    \Ecal=\frac{E}{L}, \quad \Jcal=\frac{J}{L^2}, \quad
    \Bcal=\frac{\bt}{8\pi L}
    \label{eq:S4.4}
    \eneq
\vspace*{2mm}
(i) Schwarzschild Black Holes.

Consider first a Schwarzschild black hole of mass $M^*_{bh}$ in equilibrium
with radiation enclosed in a spherical cavity with fixed radius $L$. At
sufficiently low temperature (high $\bt$ or $\Bcal$), when most of the
mass-energy is in the black hole, the total energy of the black hole and the
radiation is, to a good approximation (see the Appendix) given by the sum of
mass-energies as in flat spacetime (Hawking 1976); in Planck units
   \beeq 
   E=\Ebh+\Erad=\frac{\bt}{8\pi}+\frac{\pi^2}{15}nV\bt^{-4}
   \label{eq:S4.5}
   \eneq
where $n$ is the sum over the helicity states of the massless fields.
Translated into our classical units, (\ref{eq:S4.5}) becomes
   \beeq 
   \Ecal=\EbhC+\EradC=\Bcal+\sg L^{-2}\Bcal^{-4}, \quad
   \sg=\frac{4\pi^3}{45}\Lprth{\frac{1}{8\pi}}^4n \simeq \NV{6.91}{-6}n.
   \label{eq:S4.6}
   \eneq
 From now on we shall take $n=1$. The function $\Ecal(b)$ becomes eventful when
$\EbhC$ and $\EradC$ are of the same order of magnitude. Indeed the lower
bound for $\Ecal$ is reached when
    \beeq 
    \dr{\Ecal}{\Bcal}=0
    \label{eq:S4.7}
    \eneq
at which point (C in figure 1a) one has
    \beeq 
    \EbhC= 4 \EradC, \quad
    \Ecal=\Ecal_C=\frac54\,\Bcal_C, \quad
    \Bcal_C=(4 \sg L^{-2})^{\frac15} =0.123\,L^{-\frac25}.
    \label{eq:S4.8}
    \eneq

When $\Ecal> \Ecal_C$, the system's energy is dominated either by the black
hole or by the radiation. Indeed for $b>b_C$, when $\Ecal> 1.6\Ecal_C$, less
than 1\% of the energy is in the radiation. For $b<b_C$, when $\Ecal> 25
\Ecal_C$, less than 1\% is in the black hole mass. At higher energies, along
this later series, when $b \approx L^{-1/2}$, one has to take into account
the radiation self-gravitating effects and solve Einstein equations [Klein
(1947) in the case of pure radiation, see also the Appendix]. One finds that
there is a maximal temperature where $\Bcal_D\simeq 0.11 L^{-1/2}$ and $\Ecal_D
\simeq 0.123$ (see figure 1b). For still higher energies, $\Bcal$ increases
again till one finds a maximal energy $\Ecal_A \simeq 0.246$ where $\Bcal_A
\simeq 0.135L^{-1/2}$ beyond which the are no more equilibrium configurations.

Both curves, $\Bcal(\Ecal)$ and $\Bcal(\EradC)$ are drawn in figure 1a and 1b.
In figure 1b the two lines are almost on top of each other for $ \Ecal \go
\Ecal_D$ (due to the small mass of the hole: $\Ecal_{bh}/\EradC \approx b
\approx L^{-1/2}$) and they spiral inwards counterclockwise with an almost
common limit point Z at $\Ecal_Z=3/14$ and $\Bcal_Z \simeq 1.23 b_D\simeq
0.132 L^{-1/2}$.

Such counterclockwise spirals appear in Newtonian theory as well as in general
relativity for equations of state of the form $P=K\rho$ where $P$ is the
pressure, $\rho$ the density and $K$ a number (Chandrasekhar 1972); here
$K=1/3$. The inward spiral will not differ very much whether we keep $L$ or
the {\it proper} volume of the sphere fixed. Therefore, stability limits will
not differ significantly either.

The two linear series $\Bcal(\Ecal)$ and $\Bcal(\EradC)$ coexist between two
vertical lines for $\Ecal_C\leq \Ecal\leq \Ecal_A$. Thus between these lines
there will be always one metastable state. We emphasis the origin of $\Ecal_C
<\Ecal_A$ which is due to the different nature of the instability: with or
without a black hole at the Hawking temperature. This leads to the
scaling
   \beeq  
   \frac{E_C}{E_A} = 0.65 L^{-2/5}.
   \eneq
Very important also is the fact that there is a maximal temperature $T_D\simeq
0.37 L^{-1/2}$. Thus for $L>10^6$, $T$ is always at least three orders of
magnitude below the Planck temperature. We may thus safely neglect quantum
matter effects (York 1985), quantum gravity and spontaneous nucleation (Piran
and Wald 1984). For smaller boxes, not only should one take into account the
quantum matter effects induced by the presence of the hot black hole, but one
has presumably to abandon the thermodynamic analysis all together since the
fluctuations become important and are not governed anymore by their
thermodynamics estimates. Indeed, for those configurations, the thermal and
the dynamical characteristic times come to coincide.

Finally, we mention that Balbinot and Barletta (1988) computed the change of
the black hole entropy $\Sbh$ due to some quantum matter back-reaction. This
provides a model for black hole remnants since the tiny black hole may now be
in equilibrium with radiation for arbitrary large (or small) temperatures. One
easily shows that for cavities with $L>10^6$ one finds the linear series to be
almost unmodified and thus stability limits unaffected (contrary to what was
suggested) since the modifications are dimensionalized by the Planck mass and
since $T\lo 0.37$ $10^{-3}$. \\[2mm]
(ii) Rotating Black Holes.

Consider now a Kerr black hole at mid-height on the $z$ axis of a cylindrical
cavity of radius $L^*$. When thermodynamical equilibrium is achieved, the
radiation has the same temperature as the hole and rotates with the same
angular velocity: $\Om_{bh}=\Om_{rad}$. The peripheral velocity of the
cavity, in units of $c$, is
   \beeq 
   \nu =\frac{\Om_{rad} L^*}{c}, \quad 0\leq \nu\leq 1.
   \label{eq:S4.9}
   \eneq
Schumacher \etal\ (1992) have calculated what becomes of \EQ\ (\ref{eq:S4.6})
when the gravitational pull of the hole and the self-gravitational effects of
the radiation are neglected. This amounts to add to the energy of a rotating
black hole $\EbhC(\Bcal,\nu)$ the energy of the radiation calculated (in
special relativity) in rotating coordinates. We shall push the analysis of
that approximation beyond its presumed limits of validity in order to gain
some insight of what may still happen at very high angular momentum $(\nu\to
1)$ when general relativity has to be taken into account.

Introducing a non-dimensional parameter $h\equiv J/M_{bh}r_H$ where $r_H$
is the radius of the horizon (Okamoto and Kaburaki 1992), the mass-energy,
the angular momentum and the angular velocity of
the hole can be written
   \beeq 
   \EbhC=\Bcal(1-h^2), \quad \Jcal_{bh}=
\frac{2h}{1+h^2}(1-h^2)^2\Bcal^2 ,
   \quad 0\leq h\leq 1
   \label{eq:S4.10}
   \eneq
   \beeq 
   \Om_{bh}=\frac{4\pi}{\bt}\frac{h}{1-h^2}.
   \label{eq:S4.11} 
   \eneq
At equilibrium, $\Om_{bh}=\Om_{cav}$, thus \EQ s (\ref{eq:S4.9}) and
(\ref{eq:S4.11}) give
   \beeq 
   \nu(h,\Bcal)=\frac{1}{2\Bcal}\frac{h}{1-h^2}.
   \label{eq:S4.12}
   \eneq
For a cylinder of height $H^*=2L^*$, Schumacher \etal\ found
   \beeq 
   \Ecal(\Bcal,h)=\Bcal(1-h^2)
         +\frac32\sg L^{-2}\Bcal^{-4}\frac{1-\nu^2/3}{(1-\nu^2)^2}
   \label{eq:S4.13}
   \eneq
and
   \beeq 
   \Jcal(\Bcal,h)=\frac{2h}{1+h^2}(1-h^2)^2\Bcal^2
              +\sg L^{-2}\Bcal^{-4}\frac{\nu}{(1-\nu^2)^2}.
   \label{eq:S4.14} \eneq
Equations (\ref{eq:S4.14}) with (\ref{eq:S4.12}) immediately indicate that
most of the angular momentum is in the radiation except when $\Bcal$ is close
to $1/2$, \ie when the black hole fill up the entire cavity. For a given value
of $\Jcal$ held fixed, equation (\ref{eq:S4.14}) gives $h(\Bcal)$ which,
substituted in \EQ\ (\ref{eq:S4.13}) gives $\Bcal(\Ecal)$. The $\Bcal(\Ecal)$
function can only be written in parametric form and we have to solve it
numerically. We have drawn $\Bcal(\Ecal)$ for $\Jcal=1/40$ in figure 2a and
for $1/8$ in figure 2b.

A common feature of the two linear series of figures 1 and 2 is that black
holes and radiation coexist to the right hand side of a vertical line
$CC_{rad}$ since there is always a minimum energy $\Ecal_C (\Jcal)$ required
to find a black hole in equilibrium with the radiation. There is no maximal
energy, the equivalent of $\Ecal_A$, on the right hand side because we have not
take into account the self-gravity of the rotating radiation. There are also
two important modifications in the behavior of $\Bcal(\Ecal)$ at fixed $\Jcal$
when one compares linear series at different $\Jcal$. Both are shown in
figure 3. First, the turning point C moves in the plane $(\Ecal,\Bcal)$;
$\Ecal_C$ increases with $\Jcal$. Second, the energy of the black hole at $C$,
$\Ecal_{bh\,C}(\Jcal)$\,---\,the energy gained by the radiation between $C$
and $C_{rad}$\,---\,divided by the total energy $\Ecal_C (\Jcal)$ decreases
with $\Jcal$.

Another novelty appears for $\Jcal\go 1/40$. We remind the reader that in the
absence of radiation, it was noted that for $J_{bh}/M^2_{bh}\go 0.68$ the slope
$\pl\bt/\pl M_{bh}$ becomes negative, going through 0 (Davies 1977). When the
radiation is taken into account, there is no change of sign in $\pl\bt/\pl E$
since the radiation dominates the equilibrium configuration before one
approaches the critical ratio $0.68$ of $J_{bh}/M^2_{bh}$. For $\Jcal\go 1/40$
\fnt{                  
This limit is obtained by substituting $\nu$ from (\ref{eq:S4.12}) into
(\ref{eq:S4.14}). For a given $\Jcal$, one finds 3 values of $0\leq h\leq 1$
corresponding to 3 points on $b(\Ecal)$. Two of the points are on the left and
right of the local maximum, point $X$ in figure 2b, when $\Jcal$ is high
enough, say $1/8$. As $\Jcal$ decreases the 2 points come closer to each other
and point $X$ goes down the line $b(\EbhC)$. At about $\Jcal\simeq 1/40$ the
points coincide and $X$ is at the bottom of $b(\EbhC)$. For $\Jcal\lo 1/40$,
there is no local maximum anymore.} 
one does recover this phenomenon in a small interval of energy (see figure
2b). There are now two changes of sign upon decreasing the energy, before the
black
hole starts to evaporate. At lower energies the radiation dominates the
equilibrium again and $\pl\bt/\pl E$ returns positive. \\

\befl
{\bf 5.\ Fluctuations near the turning points  } \\[2mm]
\enfl
We shall now compute the mean square fluctuations of the rescaled variable
$\tlu$, \EQ\ (\ref{eq:S3.32}), near the turning points C and A. We shall also
find the probabilities $dW$, \EQ\ (\ref{eq:S3.142}), that the fluctuations be
big enough for the system to jump from metastable to stable configurations.
Finally we shall say a few words about characteristic times and for the
probability rates.

At the turning point $C_V=0$. Thus, for an equilibrium configuration $P$ in
the vicinity of the turning point, to lowest order in $(\bt_P -\bt_0)$ where
$\bt_0$ stands for either $\bt_C$ or $\bt_A$ one has
   \beeq 
   -C_V=\bt^2_P \Lprth{\pldr{E}{\bt}}_P
       \simeq \bt_0^2\Lprth{\pldrh{2}{E}{\bt}}_0 (\bt_P -\bt_0)
      = \Lprth{\bt^3\, \pldrh{2}{E}{\bt}}_0 \Dlu
   \label{eq:S5.4}
   \eneq
where
   \beeq 
   \Dlu=\frac{\bt_P -\bt_0}{\bt_0}, \quad |\Dlu|\ll 1.
   \label{eq:S5.5}
   \eneq
$\Dlu$ parametrises the equilibrium configurations $P$ and should not be
confused with the fluctuation out of equilibrium $\Dltlu$ defined in \EQ\
(\ref{eq:S3.141}). In terms of $\Ecal$ and $\Bcal$, one has thus
   \beeq 
   -C_V = \bt^{-2} \Lprth{\pldr{\bt}{E}}
        \simeq 8\pi L^2 \Lprth{\Bcal^3\pldrh{2}{\Ecal}{\Bcal}}_0 \Dlu.
    \label{eq:S5.6}
   \eneq
Near point C, \EQ\ (\ref{eq:S4.6}) and (\ref{eq:S4.8}) give
   \beeq 
   -(C_V)_C \simeq 1.9 L^{6/5} \Dlu .
    \label{eq:S5.7}
   \eneq
Near point A, one finds
   \beeq 
   -(C_V)_A \simeq 23 L^{3/2}(-\Dlu)
   \label{eq:S5.8}
   \eneq
(see the end of the Appendix for the numerical factor 23). The scaling law
$L^{3/2}$ itself is easily found: $b(\EradC)$ scales like $L^{-1/2}$ and the
radiation entropy scales like $b^{-3}\propto L^{3/2}$.

Near points A and C, the mean fluctuations of $\tlu$, given in \EQ\
(\ref{eq:S3.143}), are
 extremely small (and therefore the temperature fluctuations
as well) owing to the presence of the positive power of $L$ in $C_V$.
Nevertheless very close to the turning point, the
fluctuations diverge and thus will be high enough to induce a transition
towards stable configurations. We therefore estimate at which distance from
the turning points, that is, for which $\Delta u$, will the {\em mean} square
fluctuations be high enough. To this end, we note that the minimal size that
a fluctuation should possess in order to provoke the transition is $2 \Dlu$.
Indeed smaller fluctuations have not reached the minimal entropy configuration
which lies onto the unstable branch of the linear series and thus
which is as well at a distance $\Dlu$ from the turning point. Hence
the system
will most likely return towards the initial metastable configuration.
The minimal size of $\MSF{\tlu}$ is thus reached when $\MSF{\tlu}\simeq
(2\Dlu)^2$, that is when
  \beeq 
  -C_V^{-1} = 4(\Dlu)^2.
  \label{eq:S5.9}
  \eneq
Using \EQ\ (\ref{eq:S5.7}) we see that near point C this happens when $(\Dlu)_C
=0.51 L^{-2/5}$. At point A, following (\ref{eq:S5.8}), $(\Dlu)_A\simeq 0.22
L^{-1/2}$. If $L>10^6$, one finds $(\Dlu)_C< \NV{2.0}{-3}$ and $(\Dlu)_A<
\NV{2.2}{-4}$. Thus only very nearby the turning point will the mean
fluctuations
cause a phase transition. For $\Dlu$'s bigger than $(\Dlu)_C$ or $(\Dlu)_A$
the probabilities decrease drastically like $\exp(2 (\Dlu)^2 C_V))$. This
shows that when $\Dlu$ is only a few times bigger than $(\Dlu)_C$ or $(\Dlu)_A$
the metastable black hole near C and the metastable radiation near A are
perfectly stable.

Having found $dW$, we now estimate probabilities of fluctuations per unit time.
The rate at which a particular fluctuation of energy $\Dl E$ occurs is given,
by virtue of the fluctuation dissipation theorem, by the inverse time it takes
a cavity to return to equilibrium after the addition of the energy $\Dl E$.
This time depends on the peculiar dynamics of the system. For a black hole in
equilibrium with radiation, the time, following Zurek (1980), is of the order
of $\bt^4\Dl E$. For pure radiation, the time is of the order of $\bt^2\Dl E$.
Hence both times are proportional to powers of $L$. Thus probabilities per unit
time behave essentially like $dW$ itself, \ie the negative exponential of
powers of $L$ dominates completely the probability rates.  \\

\befl
{\bf 6.\  Stability and phase transitions in evolutionary sequences through
          quasi-equilibrium states: A summary }
\enfl
(i) The Microcanonical Ensembles \\[2mm]
(a) Non-rotating Black Holes and Radiation.

Consider an equilibrium configuration in the $\Ecal$-$b$ plane, say at point
F higher than point B in figure 1a. Let us remove energy by small amounts so
that the system stays practically in equilibrium and evolution takes place
along the linear series $b=b(\Ecal)$. At energies $\Ecal<\Ecal_B$, the
entropy of pure radiation $\Srad$ is greater that the entropy of the composite
system $S$. This is true only if one does not add a constant to the black hole
$S_{bh}$ which will shift the point B. We emphasis that the addition of this
constant, on the contrary, does not affect our computations of the mean
fluctuations between B and C nor the fact that the black hole will evaporate
at C. For $\Ecal<\Ecal_B$, the temperature is higher than the temperature at
$\Ecal_B$, the black hole is thus superheated (Gibbons and Perry 1978). We
have just shown in section 5 that the probability for a fluctuation to lead to
the total evaporation is completely negligible as long as $\Dlu\go(\Dlu)_C
\simeq 0.51 L^{-2/5}$. Thus the system will evolve almost down to $\Ecal_C$
staying in these superheated states. If one remove energy below $\Ecal_C$, the
black hole cannot survive in equilibrium anymore and will evaporate into
radiation to a $b=b_{C\,rad}=0.082L^{-2/5}$, that is, with a temperature given
by $5^{1/4}T_C$, since $4/5$ of the energy was in the black hole. Removing
more energy will simply cool the radiation down.

Now if one reverses the process and starts to add energy to the cavity, say
from point G in figure 1a, the evolution takes place along the linear series
$\Bcal(\EradC)$. The radiation heats up and reaches the point $B_{rad}$ where
$\Ecal_{B\,rad}=\Ecal_B$. Beyond this point the radiation finds itself also in
a superheated state. The chances to form a black hole at those energies are
exponentially small (see also Piran and Wald 1982). The radiation will thus
continue to evolve (from figure 1a to figure 1b) in that superheated state as
long as $|\Dlu|\go |\Dlu|_A\simeq 0.22 L^{-1/2}$, that is, almost up to point
A. It will never become supercooled because $b_A\simeq 0.14\,L^{-1/2}<b_{B\,
rad}\simeq 0.20\,L^{-2/5}$ for $L>10^6$. Near point A, the radiation will
collapse and form, near point $A_{bh}$,
 a black hole in equilibrium with the left over
radiation.

One has thus a closed circuit of equilibrium configurations which can be
experienced counterclockwisely only. \\[2mm]
(b) Black-Holes in Rotating Cavities\,---\,Self-Gravity Neglected.

If we compare figures 1 and 2, we see that rotation does not modify very much
the above picture. Even for fast rotation, significant differences occur only
in the late stages of evolution before evaporation. As one approaches point C
in figure 2, the black hole has already lost most of its energy (see figure
3). The higher the angular momentum, the smaller this energy. In evolutionary
sequences with decreasing energy of fast rotating cavities, evaporation of
black holes will go almost unnoticed, the first order phase transition being
very mild. Since we have not taken into account the self-gravity of the
radiation, one does not find the equivalent of the point A nor the
closed circuit.\\[2mm]
(ii) Canonical  Ensembles \\[1mm]
(a) Non-rotating Black Hole in a Cavity

If one assume that one may control the temperature at infinity instead of
letting the system be isolated, stability conditions can be read from the
same $\Bcal(\Ecal)$ diagram but rotated $90^{\circ}$ clockwise (or
equivalently one looks for horizontal tangents). Looking at figure 1, we
observe the following situation. At low temperature, high $b$, with very
little radiation, the lonely black hole cannot be stable since it has
negative heat capacity. Then the $-\Ecal(\Bcal)$ curve rotates
counterclockwise and the sequence of inwards spiraling configurations will
become more unstable each time one encounters a vertical tangent. Thus black
holes in a cavity with radiation at fixed $\Bcal$ are always unstable. We
mention that York (1985) has considered cavities with fixed temperature at the
boundary rather than fixed temperature as measured from infinity (see \EQ\
(A.3) in the Appendix). This leads to a different Legendre transformation
which defines another free energy. Therefore, it is not surprising that this
other ensemble has different stability limits when the energy approaches the
\Schw\ limit.

One should, however, question the physical relevance of canonical
configurations in general wherein self-gravitating effects are important. This
is because one cannot ignore the self-gravity of the reservoir needed to fix
the temperature. Indeed, in order to maintain properly the temperature fixed,
the reservoir has to be large compared to the system itself. Thus the reservoir
would be within its own \Schw\ radius. It appears therefore that the \Cnn\
situations have hardly any physical relevance.

Pure radiation behaves quite differently. Black body radiation in a cavity at
low temperature (and hence small mass) is stable. If we slowly heat the
cavity, the sequence of evolution is the $-\EradC(\Bcal)$ curve of figure 1.
Equilibrium configurations are all stable up to a temperature $T_D$ or down
to point $D$ where $C_V$ becomes negative. For $b$ below $b_D$ there is no
equilibrium configuration and at higher energies the configurations are all
unstable, since $-\EradC(\Bcal)$ spirals inwards counterclockwisely.  \\[2mm]
(b) Black Holes in Rotating Cavities\,---\,Self-Gravity Neglected

Black holes in rotating cavities and in a heat bath are as dull as
non-rotating black holes. They are all unstable. The situation changes,
however, somewhat when $\Jcal\go 1/40$. Then, as can be seen on figure 2b
rotated clockwise $90^{\circ}$, two vertical tangents appear at point $X$ and
$Y$ enclosing a narrow range of energies. At high $b$, \ie $b>b_X$, the \Cnn\
ensemble is certainly unstable, the specific heat being negative. Since we know
that the \MC\ ensemble is stable for those configurations, the number of
negative \Pc\ coefficients is only one (see point (d) of section 2). From the
point X, the system will stay stable if we slowly change the temperature as to
decreasing the energy from $\Ecal_X$ to $\Ecal_Y$. At point $Y$ we reach again
a vertical tangent. The counterclockwise turn of the line signals that
instability is back. All equilibrium configurations at higher temperatures are
thus unstable. Notice that the ``stabilizing" effect of the angular momentum
appears in a domain where general relativity begins to be important. \\

\befl
{\bf Acknowledgements} \\[2mm]
\enfl

We are grateful to Y.\,Okuta for providing us with figures and numerical
values, all of which are part of a work in progress. One of us R.P.\ would
like to thank J. Bekenstein and R. Brout for many clarifying discussions. \\

\befl
{\bf Appendix A: On self-gravitating radiation in a box} \\[2mm]
\enfl

The properties of global equilibrium configurations of a self-gravitating,
finite-sized, non-rotating, spherical symmetric fluid with {\it no black hole
in the center} has been studied for various reasons by Klein (1947),
Chandrasekhar (1972), Sorkin \etal\ (1982) and Page (1992). The equation of
state of a gas of photons $(n=1)$ is
   \beqn
   P=\frac13 \rho
   \eeqn{A.1}
where $\rho$ is the mass-energy density. Under local thermodynamic
equilibrium conditions [see \EQ\ (\ref{eq:S4.6})], $\rho$ is given by
   \beqn
   \rho=\fracdps{\pi^2}{15} T^4(r)
   \eeqn{A.2}
in Planck units where $T(r)$ is the local temperature function of $r$,
$T(r)=T g_{00}(r) ^{-1/2}$ where $g_{00}^{1/2}$ is the $r$-dependent lapse
function which relates the local proper time to the asymptotic one. In
particular, on the surface of the cavity, one has
   \beqn
   \quad   T(L)\sqrt{1-\fracdps{2\Erad(L)}{L}}=T(L)\sqrt{1-2\EradC(L)}=
   T=\fracdps{1}{\bt}.
   \eeqn{A.3}
where $T$ is the ``temperature as measured at infinity". The distribution
$\rho(r)$ is given by the Tolman-Oppenheimer-Volkoff equation which can be
solved as follows (Sorkin \etal\ 1981). Set
   \beqn
   q=4\pi r^2\rho, \quad
   \mu=\fracdps{4\pi}{r} \dps\int^r_0 dr'r'^2\rho'.
   \eeqn{A.4}
Then the TOV equation reads
   \beqn
   \drdps{q}{\mu}=\fracdps{2q(1-4\mu-(2/3)q)}{(1-2\mu)(q-\mu)}.
   \eeqn{A.5}
Initial conditions are $\mu=q=0$, this leads to a family of solutions
parametrized by the density at $r=0$. On the boundary $r=L$
   \beqn
   q(L)=4\pi L^2\rho(L), \quad \mu(L)=\EradC.
   \eeqn{A.6}
The second equation is used to express the density at $r=0$ in terms of
$\EradC$. Thus, the solution of \EQ\ (A.5) defines, with (A.3) and (A.6) the
function $\Bcal(\EradC)$
   \beqn
   \Bcal^4=\fracdps{3\sg L^{-2}}{q(1-2\EradC)^2}
\eeqn{A.8}
where $q$ is a function of $\EradC$.

When self gravity is negligible ($q\simeq 3\mu \ll 1$),  $\Bcal(\EradC)$
reduces to the expression valid in flat space time $E_{rad}=\sigma VT^4$.
When $\EradC>0.1$, self gravity is important and $\EradC(\Bcal)$ is given by
(A.8). Special points are given in the table (A.9); the limit point Z of
figure 1 is where the density at $r=0$ is infinite. \\

\begin{tabular}{llllll}\hline\hline
Special points$^*$           &  D       & A       & E       & Z \\
$\EradC \simeq$              &  $0.123$ & $0.246$ & $0.235$ & $0.214$ \\
$\Bcal\simeq L^{-1/2}\times$ &  $0.107$ & $0.135$ & $0.140$ & $0.132$ \\
$T\simeq   L^{-1/2}\times$ &  $0.372$ & $0.295$ & $0.284$ & $0.301$ \\ \hline
\end{tabular}
\hfill (A.9)

$^*$See figure 1. For comparison,
 $\Ecal_{C}\simeq 0.154\,L^{-2/5}$ and
$b_{C_{rad}}\simeq 0.082L^{-2/5}$.

\vspace*{2mm}
When a black hole is present in the center of the cavity, (A.1) and (A.2) are
no more valid. One has to take into account the quantum behavior of the mass
energy density since the Tolman relations certainly break down near the
horizon. The quantum version of the density is provided by the expectation
value of the time-time component of the stress energy tensor in the so-called
Hartle-Hawking ``vacuum" (Howard 1984). One finds first that the contribution
to the gravitational mass for $2M_{bh}<r<6M_{bh}$ is of the order of
$M_{bh}^{-1}$ (in our units, it means $b^{-1}$) and secondly that for
$r>6M_{bh}$ one may approximate the mass energy density by (A.2). Thus one may
approximate (\ref{eq:S2.3}) by
   \beqn
   m(r,M_{bh})=M_{bh}+\int^r_{6M_{bh}}dr'r'^2
                   M_{bh}^4 (1-2M_{bh}/r')^{-2}.
   \eeqn{A.10}
(again for cavities with $L>10^6$). We have also neglected in (A.10) the
self-gravity of the radiation. For $\Bcal>\Bcal_C$, this is certainly valid
since $E_{rad}<E/5$ and since $E_C\ll L/2$. For smaller black holes ($E_{bh} <
(4/5)E$), the configurations are unstable and it is therefore useless to
calculate the corrections. From (A.10) and for $\Bcal>\Bcal_C$, it is easy to
verify that the corrections to $\Bcal(\Ecal)$ are small and have no effect on
thermodynamic stability limits. The reader might consult Page (1992) to find
an explicit evaluation of corrections to the entropy.

We now give the derivation of $(C_V)_A$ appearing in \EQ\ (\ref{eq:S5.8}). We
use \EQ\ (A.8) with \EQ\ (A.6) and obtain the following expression for
$d\Ecal/d\Bcal$ (we now drop the index $rad$)
    \beqn
    \fracdps{3\sg L^{-2}}{4\Bcal^3}\drdps{\Ecal}{\Bcal}
    =\fracdps{q(1-2\Ecal)^3(q-\Ecal)}{2(\frac83 q-1+2\Ecal)}.
    \eeqn{A.11}
Using again (A.6) with (\ref{eq:S5.5}), we also obtain for $dq/d\Bcal$
    \beqn
    \fracdps{3\sg L^{-2}}{4\Bcal^3}\drdps{q}{\Bcal}
    =\fracdps{q^2(1-2\Ecal)^2(1-4\Ecal-\frac23 q)}
             {(\frac83 q-1+2\Ecal)}.
    \eeqn{A.12}
These two equations give us a means to calculate the second derivative of
$d^2\Ecal/d\Bcal^2$
    \beqn
    \Lprth{\fracdps{3\sg L^{-2}}{4\Bcal^3}\drhdps{2}{q}{\Bcal}}
    =\pldrdps{ }{q}
     \Lbr{\fracdps{q(1-2\Ecal)^3(q-\Ecal)}{2(\frac83 q-1+2\Ecal)}}
     \Lprth{\drdps{q}{\Bcal}}_A.
     \eeqn{A.13}
At point A, the first derivative is zero
   \beqn
   (-C_V)_A =64\pi L^2\Bcal_A \fracdps{\Ecal_A(1-2\Ecal_A)}
                                  {1-\frac{14}{3}\Ecal_A}\Dlu.
   \eeqn{A.14}
With $b_A$ given in the above table, one has thus
   \beqn
   (-C_V)_A \simeq 23\,L^{3/2}(-\Dlu).
   \eeqn{A.15}

\newpage
\befl
{\bf References}
\enfl
\ssk
Balbinot R and Barletta A 1989 \CQG \vol{195} 203 \\
Brown J D, Comer G L, Melmed J, Martinez E A, Whiting B F and York J W 1990
\CQG\ \vol{7} 1433 \\
Callen H B 1985 {\it Thermodynamics and introduction to thermostatics}, p.425
(Wiley: New York) \\
Carlip S and Teitelboim G 1993 {\it The Off-shell Black Hole} gr-qc 9312002 \\
Chandrasekhar S 1972 in {\it General Relativy}  ed L O'Raifeartaigh (Clarendon
Press) \\
Cocke W J 1965 {\it Ann Inst Henri Poincar\'e} 283 \\
Davies P C W 1977 {\em Proc.\,Roy.\,Soc.\,Lond.} \vol{A353} 499 \\
Gibbons G W and Hawking S W 1977 \PRD\ \vol{15} 2752\\
Gibbons G W and Perry M J 1978 {\it Proc. Roy. Soc. Lond.} \vol{A 358} 467\\
Hawking S W 1976 \PRD\ \vol{13} 191 \\
Hawking S W 1978 \PRD\ \vol{18} 1747 \\
Horwitz G and Katz J 1978 \ApJ\ \vol{222} 941  \\
Horwitz G and Weil D 1982 {\it Phys. Rev. Let.} \vol{48} 219\\
Howard K W 1984 \PRD\ \vol{30} 2532\\
Kaburaki O, Okamoto I and Katz J 1993 \PRD\ \vol{47} 2234 \\
Katz J 1978 \MN\ \vol{183} 765 \\
---------\,1979 \MN\ \vol{189} 817 \\
Katz J and Manor Y 1975 \PRD\ \vol{12} 956  \\
Katz J, Okamoto I and Kaburaki O 1993 \CQG\ \vol{10} 1323 \\
Klein O 1947 {{\it Ark.\,Mat.\,Astr.\,Fys.\,A}} \vol{34} 1 \\
Landau  L D and Lifshitz E M 1980 {\it Statistical physics} 3rd ed (Pergamon:
Oxford) \\
Landsberg P T 1990 {\it Thermodynamics and statistical mechanics} (Dover) \\
Ledoux P 1958 {\it Stellar stability} in Handbuch der Physik ed Flugge S
\vol{51} \\
Lynden-Bell D and Wood R 1968 \MN\ \vol{138} 495 \\
Lyttleton R A 1953 {\it Theory of rotating fluid masses} (Cambridgr University
Press)  \\
Okamoto I and Kaburaki O 1990 \MN\ \vol{247} 244 \\
Okamoto I, Katz J and Parentani R 1994 ``A Comment on Fluctuations and
Stability with Application to Superheated Black Holes" Preprint \\
Page D N 1992 in {\it Black hole physics} eds De Sabatta V and Zhang Z (Kluwer
Holland) \\
Parentani R 1994 ``The Inequivalence of Thermodynamic Ensembles" Preprint \\
Piran T and Wald R M 1982 {\it Phys Let A} \vol{90} 20 \\
Regge T and Teitelboim G 1974 {\it Ann. Phys.} \vol{88} 286\\
Schumacher B, Miller W A and Zurek W H 1992 \PRD\ \vol{46} 1416. \\
Sorkin R D, Wald R M and Zhang Z J 1982 {\it Gen Rel Grav} \vol{13} 1127 \\
Thompson J M T 1979 {\it Phil Trans R Soc Lond} \vol{292} 1386. \\
Tolman C 1934  {\it Relativity, thermodynamics and cosmology} p.318
(Clarendon) \\
York J W 1985 \PRD\ \vol{31} 775 \\
York J W 1986 \PRD\ \vol{33} 2092 \\
Zurek W H 1980 {\it Phys Let} A 77 399 \\

\newpage
\befl {\Large\bf Figure captions }\\[2mm] \enfl

\nid
{\bf Figure 1.} $b(\EbhC)$, $b(\EradC)$ and $b(\Ecal)$ are drawn for $L=10^6$,
in figure 1a for $\Ecal\lo 10^{-3}$ and in figure 1b for $\Ecal\go 10^{-1}$.
Between $\Ecal=10^{-3}$ and $10^{-1}$, $b(\EradC)$ and $b(\Ecal)$ come closer
and closer to the point that in figure 1b the two lines are indistinguishable.
The line $b(\Ecal)$ through points $FCQDAZ$ is a counterclock inward spiral.
The dotted line in the lower left hand corner of figure 1b is the
non-relativistic (no selfgravity) $b(\EradC)$ for comparison. \\

\nid
{\bf Figure 2.} $b(\EbhC,\Jcal)$, $b(\EradC,\Jcal)$ and $b(\Ecal,\Jcal)$
 curves for constant
angular momentum $\Jcal$ and $L=\NV{2.65}{4}$. The
low value of $L$ is to make clear
figures. Figure 2a is for $\Jcal=1/40$ and figure 2b is for $\Jcal=1/8$. Both
lines are drawn with the same limits of $b$ and $\Ecal$, showing the
displacement to right for increasing $\Jcal$
of the linear series $b(\EradC,\Jcal)$, $b(\Ecal,\Jcal)$
and of the point C. Once the equilibrium configurations leave the linear
series
$b(\EbhC,\Jcal)$, the black hole starts to loose mass with respect to
the radiation, see figure 3. \\

\newcommand{\EbhCC}{{\cal E}_{bh\,C}}
\nid
{\bf Figure 3.} This figure displays two lines: (a) The minimum of energy
$\Ecal_C(\Jcal)$ as a function of increasing angular momentum. The scale of
$\Ecal_C$ displayed on the left is the same as in figure 2. The curve is
parametrized in real values of $\Ecal_C$. Notice that beyond $\Jcal=3/16
\approx 0.2$, the energy $\Ecal_C$ is highly relativistic and the
non-relativistic curve is likely to be different. (b) The ratio $\EbhCC/
\Ecal_C=b_C/\Ecal_C$ with its scale displayed on the right. For $\Jcal=0$,
$\EbhCC/\Ecal_C=4/5$; this is well above the limits of the drawing.

\end{document}